\documentclass[reprint,aps,prl,twocolumn,superscriptaddress,preprintnumbers,letterpaper,natbib,floatfix]{revtex4-1}
\usepackage{amssymb}
\usepackage{amsmath}
\usepackage{epsfig}
\usepackage{comment}
\usepackage{color}
\usepackage{hyperref}
\usepackage{cleveref}
\usepackage{multirow}
\usepackage{capt-of}
\usepackage{siunitx}
\usepackage{graphicx}
\usepackage{slashed}
\usepackage{tabularx}
\usepackage{enumitem}
\usepackage{multirow}
\usepackage{booktabs}
\usepackage{isotope}
\usepackage[dvipsnames]{xcolor}
\usepackage{bm}
\usepackage{array,mathtools}

\usepackage{physics}

\newcommand{\mysec}[1]{\paragraph*{#1.}\!\!\!\!\!---}

\newcommand{\bea}{\begin{eqnarray}}
\newcommand{\eea}{\end{eqnarray}}

\newcommand{\beq}{\begin{equation}}  
\newcommand{\eeq}{\end{equation}}

\newcommand{\eqnref}[1]{Eq.~(\ref{eqn:#1})}
\newcommand{\figref}[1]{Fig.~\ref{fig:#1}}

\begin{document}
\title{A Lighter QCD Axion from Anarchy}
\author{Fatemeh Elahi} \email{felahi@uni-mainz.de}
\author{Gilly Elor}\email{gelor@uni-mainz.de}
\author{Alexey Kivel}\email{alkivel@uni-mainz.de}
\author{Julien Laux}\email{jlaux01@uni-mainz.de}
\author{Saereh Najjari}\email{snajjari@uni-mainz.de}
\author{Felix Yu}\email{yu001@uni-mainz.de}
\affiliation{\footnotesize \sl PRISMA$^+$ Cluster of Excellence \& Mainz Institute for Theoretical Physics\\
Johannes Gutenberg University, 55099 Mainz, Germany}

\begin{abstract}
We introduce the \emph{Anarchic Axion}, a class of axion models which solves the Strong CP problem within current nEDM constraints 
with a lighter than usual QCD axion, thus populating new parameter space that ongoing and future experiments target.  
The Anarchic Axion is driven light by a soft breaking of the Peccei-Quinn symmetry, which also predicts a residual neutron electric dipole moment.
We introduce a novel measure to quantify the tuning required for large deviations from the usual QCD axion band.  
In addition to motivating searches for unusually light axions,
this work establishes a new target for axion effective field theory.
\end{abstract}

\maketitle
\preprint{MITP-23-002}

The Peccei-Quinn (PQ) solution~\cite{Peccei:1977hh, Peccei:1977ur} to the Strong Charge-Parity (CP) problem of quantum chromodynamics (QCD) predicts relationships between the axion mass $m_a$, decay constant $f_a$ and axion-photon coupling $g_{a\gamma \gamma}$. In this Letter, we introduce a new class of multi-scalar axion models ---  \emph{The Anarchic Axion} ---so dubbed as to emphasize that the PQ symmetry arises accidentally. 

The $U(1)_{\rm PQ}$ symmetry is softly broken leading to a qualitative solution to the Strong CP problem that deviates from the traditional QCD axion band, importantly populating a region of parameter space targeted by many on-going experiments.   Related work to solving the strong CP problem departing from the canonical band include Refs.~\cite{Holdom:1982ex, Gherghetta:2016fhp, Agrawal:2017eqm, Agrawal:2017ksf, Agrawal:2017evu, Hook:2018jle,  Gaillard:2018xgk, DiLuzio:2021pxd, Kivel:2022emq}.  Furthermore, the new parameter space corresponds to an \emph{almost} perfect alignment between the soft breaking vacuum 
and the $\theta_{\rm QCD}$ vacuum. This provides a unique handle on the Axion Quality Problem~\cite{Holdom:1982ex, Kamionkowski:1992mf, Kim:2008hd, Darme:2021cxx, Dine:2022mjw, Banerjee:2022wzk, Bonnefoy:2022vop} allowing the introduction of a novel measure for quantifying the fine tuning.  

In this Letter we first introduce the field content and the potential of an Anarchic Axion model. Given an appropriate basis choice, we compute the $m_a$, $f_a$ and $g_{a \gamma \gamma}$ relations, and present the new parameter space where experiments can hunt for the Anarchic Axion. We then discuss the quality of this soft solution to the Strong CP problem and introduce a fine tuning measure to quantify the axion quality.  We conclude by commenting on possible natural UV completions and future directions.

\mysec{The Anarchic Axion}
The particle content and charge assignments of the model, summarized in Table.~\ref{table:field_content}, consists of three complex scalars: $H_1$ and $H_2$ which can be identified with the Higgs doublet fields in DFSZ axion constructions~\cite{DINE1981199, Zhitnitsky:1980tq}, and a gauge singlet $\Phi$. Standard Model (SM) fermions coupling to $H_1$ and $H_2$ will mediate the requisite effective operator coupling the Anarchic Axion and the anomalous $G\tilde{G}$ QCD term. 
\begin{table}[t]
\renewcommand{\arraystretch}{1.15}
  \setlength{\arrayrulewidth}{.15 mm}
\centering
\tiny
\resizebox{0.48\textwidth}{!}{
\begin{tabular}{| c||ccc|c||c |}
\hline
 Field & $SU(3)_c$ & $SU(2)_L$ & $U(1)_Y$ & $\mathbb{Z}_5$ & $U(1)_{PQ}$ \\ 
 \hline
 $Q_L^i$ & 3 & 2 & 1/6  & 0 & $X_Q$ \\
$u_R^i$ & 3 & 1 & 2/3  & 1 & $X_Q$-$X_1$ \\
$d_R^i$ & 3 & 1 & -1/3  & 0 & $X_Q$-$X_2$ \\
$L_L^i$ & 1 & 2 & -1/2  & 0 & $X_L$ \\
$e_R^i$ & 1 & 1 & -1  & 0 & $X_L$-$X_2$ \\ 
\hline
$H_1$ & 1 & 2 & -1/2  & 4 & $X_1$ \\
$H_2$ & 1 & 2 & 1/2  & 0 & $X_2$ \\ 
$\Phi$ & 1 & 1 & 0  & 1 & $X_3$ \\
\hline
\end{tabular}
}
\caption[Field content]{The field content of the Anarchic Axion model.}
\label{table:field_content}
\end{table}
The scalar potential is
\begin{subequations}
\begin{align}
\label{eqn:general_potential}
\nonumber &V=  \sum_{i=1,2} \left( \mu_i^2 |H_i|^2 + \lambda_i |H_i|^4 \right) + \lambda |H_1|^2 |H_2|^2 + \lambda' |H_1 H_2|^2  \\
& \quad + \mu_3^2 |\Phi|^2 + \lambda_3 |\Phi|^4 + \lambda_{13} |H_1|^2 |\Phi|^2 + \lambda_{23} |H_2|^2 |\Phi|^2 \,, \\
\label{eqn:general_potential_breaking} 
& V_{\rm break}^{C_\lambda}  = - C_\lambda H_1 H_2 \Phi +\text{h.c.} \,, 
\end{align}
\label{eqn:originalPotential}
\end{subequations}
\noindent which is invariant under the global  $U(1)_{H_1} \times U(1)_{H_2} \times U(1)_{\Phi}$ symmetry up to a gauge-invariant term with a coupling $C_\lambda$ of mass dimension 1.
Additional gauge symmetry preserving terms are forbidden by invoking a $\mathbb{Z}_5$ symmetry at high scales.
The $\mathbb{Z}_5$ allows a term~\eqnref{general_potential_breaking} which breaks the global symmetry down to $U(1)_Y \times U(1)_X$, where $U(1)_Y$ is identified with SM hypercharge and $U(1)_X$ will be identified with the \emph{accidentally} arising PQ symmetry, with $X_3 = -X_1 - X_2$.

We choose the parameters of the potential such that all three complex scalar fields acquire a vacuum expectation value (vev) $v_{1,2,3}$. Writing the electrically neutral fields in a non-linear representation, we have $\sqrt{2} H_i^0 = (v_i+h_i) e^{i a_i/v_i}$ and $\sqrt{2} \Phi = (v_3+h_3)e^{i a_3/v_3}$, where $h_i$ define the CP even radial modes and $a_i$ define the CP odd angular modes. 

Given an appropriate choice of basis, the angular modes can be rewritten as two Goldstone fields $a$ and $A$. 
We will then derive a basis-invariant anomalous CP-Violating (CPV) phase in the QCD sector,  providing a mass for the Goldstone modes and allowing the identification of one mode $a$ with the axion which solves the strong CP problem in the $\mathbb{Z}_5$ symmetric phase while the other mode $A$ is heavy.

The interesting phenomenology of the Anarchic Axion is the deviation from the canonical QCD axion band in $\{m_a, f_a\}$ parameter space (and similarly in the axion-photon effective coupling), which results from the introduction of soft PQ breaking at low scales. Specifically,
\bea
& V_{\rm {break}}^{B_\mu}  = - B_\mu H_1 H_2+\text{h.c.} \,, \label{eqn:general-potential-breaking-Bmu} 
\eea
which further breaks the symmetry down to  $U(1)_Y$. The $B_\mu$ coupling is of mass dimension 2 and breaks the $\mathbb{Z}_5$ symmetry explicitly. The magnitude of $B_\mu$ can be generated by a $\mathbb{Z}_5$ symmetric UV completion and would therefore be suppressed by a heavy UV scale.

We parameterize the symmetry breaking couplings as $B_\mu = |B_\mu| e^{-i \theta_\mu}$ and  $C_\lambda = |C_\lambda| e^{-i \theta_\lambda}$.  The original global symmetry can be used to render $\theta_{\lambda}$ unphysical, and hence $C_{\lambda}$ is a real parameter.

\mysec{The Goldstone basis}   
\label{subsec:Goldstonebasis}
Following the procedure discussed in the supplementary material, we perform a basis transformation to align the Goldstone fields with the $U(1)_Y \times U(1)_X$ symmetries, which isolates the Goldstone eaten by the gauging of hypercharge. In this new basis, the angular potential is 
\begin{align}
    \label{eqn:scalarPotentialNewBasis} 
   & V_\text{ang}=-|B_\mu| \Bigl[ \prod_{i=1}^2\left(v_i+h_i\right) \Bigr] \cos \left(\frac{a}{v_a}+\frac{A}{v_A} \delta^2 -\theta_\mu \right)\\
    &\qquad -\frac{|C_\lambda|}{\sqrt{2}} \Bigl[ \prod_{i=1}^3\left(v_i+h_i\right) \Bigr] \cos \left(\frac{A}{v_A}(1+\delta^2)-\theta_\lambda \right) \, , \nonumber
\end{align}
where $\delta=v_A/v_a$, $v_1v_2=vv_a/\sqrt{1+\delta^2}$ and $v_3=v_A/\sqrt{1+\delta^2}$, with $v = \sqrt{v_1^2 + v_2^2} \approx 246$~GeV is the electroweak vev.

Next, we include the correction to the potential arising from QCD instantons induced by the coupling of SM quarks to $H_1$ and $H_2$.
The effective Lagrangian in the Anarchic Axion model is
\begin{align}
\label{eqn:LQCD}
    \mathcal{L}_{G\tilde{G}} \supset &
    \frac{g_s^2}{32\pi^2} \left( \bar{\theta}_{\text{SM}} - N_g \left( \frac{a}{v_a} + \delta^2 \frac{A}{v_A} \right) \right)  G_{\mu\nu}^a \tilde{G}^{a\mu\nu} \,, 
\end{align}
where $N_g$ is the number of quark generations and $\bar{\theta}_{\text{SM}} \equiv \theta_{\text{QCD}} + \arg \det Y_u Y_d$ for the SM quark Yukawa matrices. 

By applying global $U(1)$ transformations on $H_1$, $H_2$, $\Phi$ and the SM quarks, we can reshuffle the separate phases from~\eqnref{scalarPotentialNewBasis} and~\eqnref{LQCD} into a new $\bar{\theta}$ defined via
$\bar{\theta}_{\text{SM}} - N_g \theta_\mu \equiv N_g \bar{\theta}$. 
We may choose $U(1)$ phases such that $\bar{\theta}$ is only in the $B_{\mu}$ contribution to the potential, and hence below $\Lambda_{\text{QCD}}$ the corresponding $a$ and $A$ fields experience the instanton potential,
\begin{align}
    \mathcal{L}_{G\tilde{G}} &\supset  
    \Lambda_\text{QCD}^4\cos \left( N_g\left(\frac{a}{v_a}+\delta^2\frac{A}{v_A}\right) \right) \,,
\label{eqn:LQCD2}
\end{align}
where $\Lambda_\text{QCD}^4 \equiv \frac{m_u m_d}{(m_u + m_d)^2} m_\pi^2 f_\pi^2$. Via the Peccei-Quinn mechanism, $\bar{\theta}$ is relaxed to the \emph{observable} CPV parameter $\bar{\theta}_{\text{eff}}$, seen as the tadpoles effects of $a$ and $A$ in~\eqnref{LQCD2}.
Consequently, the total angular potential for $a$ and $A$ fields is now
\begin{align}
    -V_\text{ang}
    = \,\, & \, \Lambda_{\text{QCD}}^4\cos \left(N_g \left(\alpha + \alpha' \delta^2 \right) \right) \label{eqn:VangInst} \\ 
    &+ \Lambda_{\text{QCD}}^4 \frac{v_a}{v_{\text{max}}} 
    \cos \left( \alpha + \alpha'\delta^2 + \bar{\theta} \right)  \nonumber\\
    &+\frac{|C_\lambda| v v_A^2}{\sqrt{2} \delta (1+\delta^2)} 
    \cos \left( \alpha' (1+\delta^2) \right)
    \, ,  \nonumber
\end{align}
where we have introduced $\alpha \equiv a / v_a$, $\alpha' \equiv A / v_A$  as convenient notation for the fields, and we have 
\begin{align}
    v_{\text{max}} & \equiv \frac{\Lambda_\text{QCD}^4}{|B_\mu|v}\sqrt{1+\delta^2}\,
    \label{eqn:vmaxandvmin}
\end{align}
as the extremal value of the PQ vev $v_a$.  For $v_a > v_{\text{max}}$ the hierarchy of the terms in the angular potential would flip and the field $a$ could not be used anymore to relax $\bar{\theta}_\text{eff}$.

\mysec{Relaxation and heavy $A$ mass}
\label{subsec:Relaxation}
To ensure $A$ is heavy, we will necessarily take $|C_{\lambda}| \gg |B_\mu|^{1/2},\, \Lambda_\text{QCD}$, such that the mass of $A$ arises dominantly from the $C_{\lambda}$ contribution to the potential, yielding \begin{equation}
    m_{A}^2 = \frac{|C_\lambda| v}{\sqrt{2}} \left(\frac{1}{\delta}+\delta + \mathcal{O} \left( \frac{\Lambda_\text{QCD}}{|C_\lambda|} \right) \right) \,,
\label{eqn:mA0}
\end{equation}
setting the scale of $A$ well above the electroweak scale.  The $v_{\text{max}}$ scale is the energy where the soft PQ breaking parameter must be nearly aligned to the $\bar{\theta}_{\text{SM}}$ to avoid neutron electric dipole moment (nEDM) constraints~\cite{Abel:2020pzs}. Pragmatically, as will be shown below, $v_\text{max}$ corresponds to the largest possible scale suppression in $g_{a \gamma \gamma}$ for $\bar{\theta}\approx\pi$. 

\mysec{The Anarchic Axion Parameters}
We now derive the $m_a$, $f_a$ and $g_{a\gamma \gamma}$ relations.
After spontaneous breaking of the PQ symmetry by $v_a$, 
the Goldstone bosons $a$ and $A$ acquire tadpoles $\alpha_0$ and $\alpha'_0$, respectively.
Importantly, the unphysical $\theta_{\lambda}$ also allows us to shift $\alpha'_0$ purely into $\alpha_0$, leaving $\alpha'_0$ unobservable, as seen in~\eqnref{VangInst}.
Consequently,
the tadpole $\alpha_0$ entirely generates $\bar{\theta}_{\text{eff}} = -\alpha_0$,
giving a potentially measurable nEDM.
Note that in this basis, $\alpha_0$ 
entirely captures the deviation from the canonical DFSZ due to non-vanishing $B_\mu$.

Expanding~\eqnref{VangInst} about the minimum and dropping constant terms and the heavy $A$ field, yields
\begin{align}
 &   -\frac{V_\text{ang}}{\Lambda_\text{QCD}^4 } \, = \alpha \left(N_g \sin(-N_g\bar{\theta}_\text{eff}) +\frac{v_a}{v_{\text{max}}} \sin(\bar{\theta}-\bar{\theta}_\text{eff}) \right) \nonumber \\
     &\quad + \,\,\frac{1}{2} \alpha^2  \left(
    N_g^2 \cos(N_g\bar{\theta}_\text{eff})  +\frac{v_a}{v_{\text{max}}} \cos (\bar{\theta}-\bar{\theta}_\text{eff} ) \right)\,. 
\label{eqn:Vang_exp}
\end{align}
For $|C_{\lambda}| \gg |B_\mu|^{1/2},\, \Lambda_\text{QCD}$, $a$ 
is already in its mass basis, where 
$m_a$ is given up to $\mathcal{O}(\Lambda_\text{QCD}/|C_\lambda|)$ corrections by
\begin{align}
\label{eqn:mamA}
    m_a^2&=
\frac{\Lambda_\text{QCD}^4}{v_a^2}\left(N_g^2\cos\left(N_g\bar{\theta}_\text{eff}\right)+\frac{v_a}{v_{\text{max}}}\cos\left(\bar{\theta}-\bar{\theta}_\text{eff}\right)\right) \ .   
\end{align}
Using~\eqnref{mamA}, the axion decay constant is 
\begin{align}
\label{eqn:fa}
\frac{1}{f_a} &\equiv \frac{N_g}{v_a} = -\frac{ \cos(\bar{\theta} - \bar{\theta}_\text{eff}) }{ 2 N_gv_{\text{max}} \cos( N_g \bar{\theta}_\text{eff})} \\ \nonumber 
&+\, \sqrt{\frac{m_a^2}{\Lambda_\text{QCD}^4 \cos(N_g \bar{\theta}_\text{eff})} + \left( \frac{\cos(\bar{\theta} - \bar{\theta}_\text{eff})}{ 2 N_g v_{\text{max}} \cos(N_g \bar{\theta}_\text{eff})} \right)^2} \ .
\end{align}
Note that in the limit $v_a\ll v_{\text{max}}$, we recover the canonical relation $m_a^2 f_a^2 = \Lambda_\text{QCD}^4$. 

Finally, to evaluate the axion-diphoton coupling, we partition the irreducible $U(1)_{\text{em}}$ anomaly shared by $a_1$ and $a_2$ into the mass eigenstate $a$, giving 
\begin{align}
\label{eqn:Gagg}
    g_{a\gamma\gamma}=&\frac{e^2}{8\pi^2}\left(\frac{E}{N}-1.92\right)\frac{1}{f_a}\nonumber\\
    &\left(1-\frac{m_a^2}{\sqrt{2}v|C_\lambda|} + \mathcal{O} \left( \frac{\Lambda_\text{QCD}^2}{|C_\lambda|^2} \right) \right) \, ,
\end{align}
where $E/N = 8/3$, analogous to the DFSZ case~\cite{ParticleDataGroup:2022pth}.

\mysec{CP Violation and nEDM}
\begin{figure}[t!]
    \centering
    \includegraphics[width=0.48\textwidth]{
    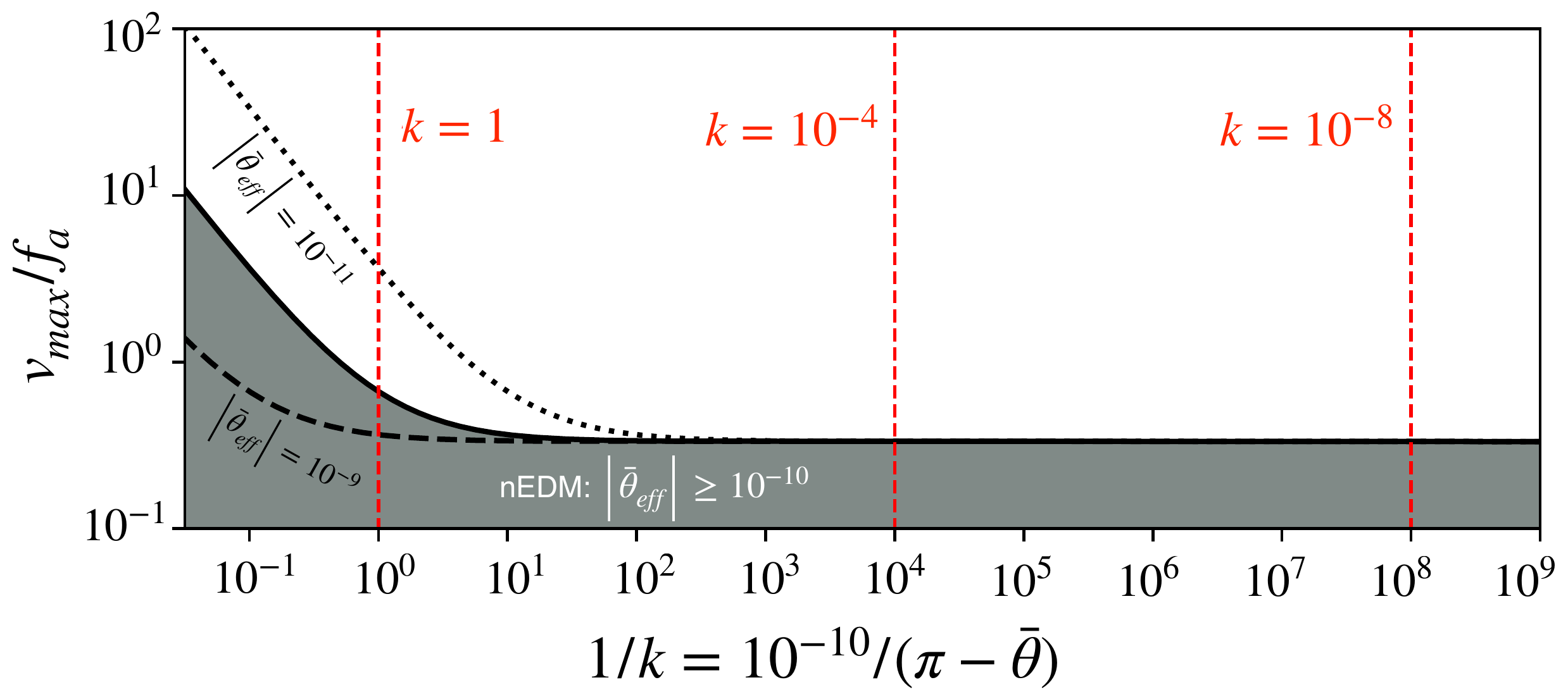}
    \caption{Contours correspond to values of  physical CPV $|\bar{\theta}_{\rm eff}|$, obtained from~\eqnref{thetaeff}. The gray region corresponds to parameter space constrained by the nEDM measurement. 
    }
    \label{fig:nEDMEllipse}
\end{figure}
We now return to the derivation of the tadpole $\alpha_0$ acquired by $a$, and the resulting observable CPV $\bar{\theta}_{\text{eff}}$ constrained by measurements of the nEDM.  To make contact with phenomenology, it will be useful to consider the leading order contribution to $\bar{\theta}_{\rm eff}$. 
The first term of~\eqnref{Vang_exp} encodes the residual CPV $\bar{\theta}_\text{eff}$ in the Anarchic Axion model, where the leading contribution up to 
$\mathcal{O}((\pi - \bar{\theta})^2)$ is given by
\begin{equation}
\label{eqn:thetaeff}
    \bar{\theta}_\text{eff}= \frac{2(\pi-\bar{\theta})}{-1+\sqrt{1+\frac{4N_g^2m_a^2v_\text{max}^2}{\Lambda_\text{QCD}^4}}} \, .
\end{equation}
Contours of $\bar{\theta}_{\rm eff}$ are show in Fig.~\ref{fig:nEDMEllipse}. We define $k \equiv (\pi - \bar{\theta})/10^{-10}$ to capture the sensitivity to the deviation of $\bar{\theta}$ from $\pi$, {\it i.e.} for $k \lesssim 1$ a tuning will be required and we saturate at
$v_{\text{max}} / f_a \to 1 / N_g$. 
The white region is allowed by the nEDM bound. For $k \gtrsim 1$, we recover the DFSZ solution to strong CP, relaxing the required tuning.

%
\mysec{Results}
\begin{figure*}[t!]
\includegraphics[width=0.95\textwidth]{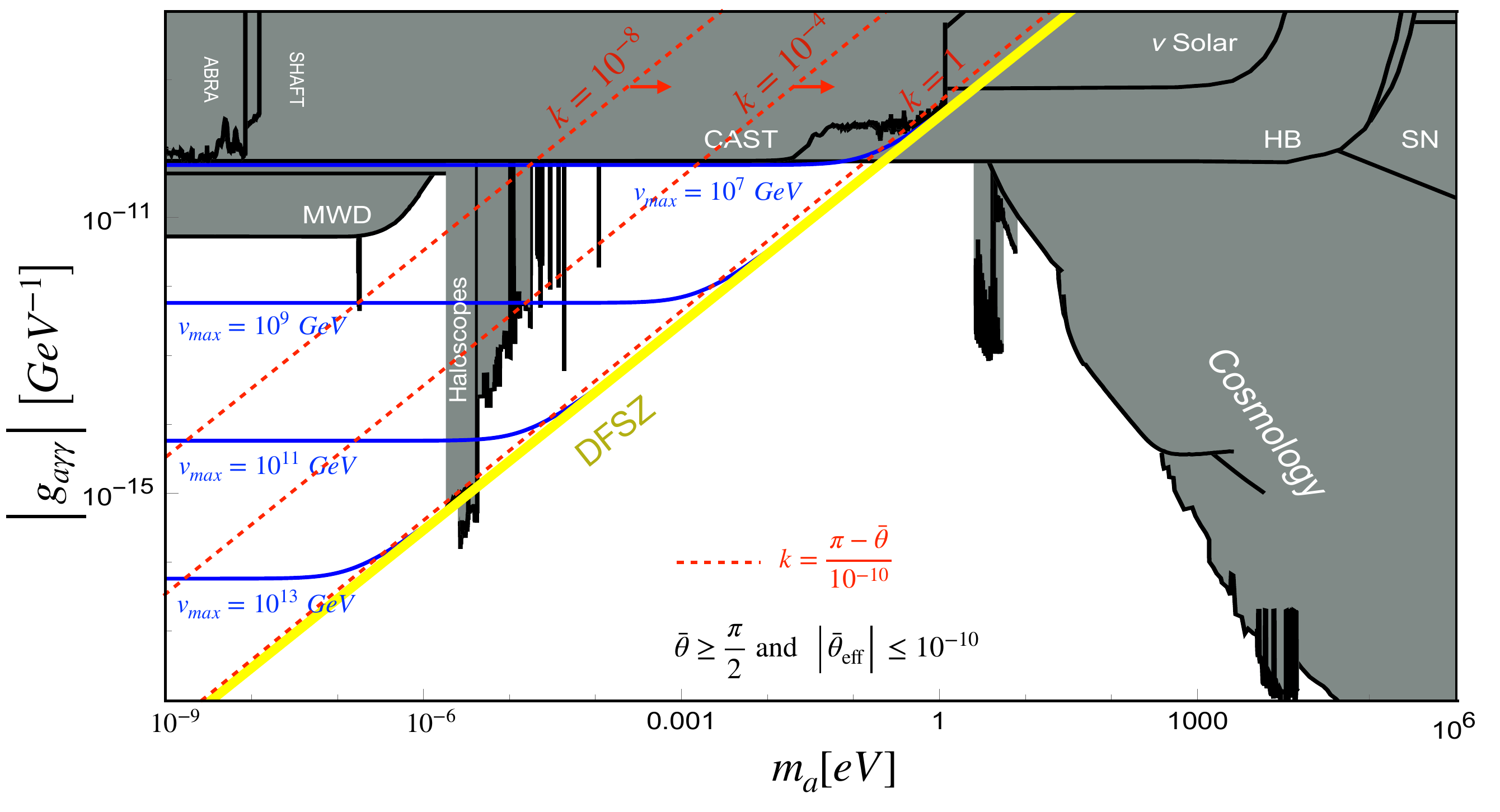}
\caption{Parameter space for the axion-diphoton coupling in the Anarchic Axion model consistent with current nEDM constraint~\cite{Abel:2020pzs}.  Experimental limits, shown by the gray shaded regions are extracted from Ref.~\cite{AxionLimits}. We show representative values of $v_{\text{max}}$ and $k$ that highlight the accessible light axion parameter space probed by ongoing haloscope and microwave cavity experiments.}
\label{fig:gagg}
\end{figure*}
In~\figref{gagg}, we display $g_{a\gamma\gamma}$ vs. $m_a$ for the Anarchic Axion. The DFSZ axion line is shown in yellow. Regions targeted by experimental searches or constrained by astrophysical considerations are shaded out in gray~\cite{ParticleDataGroup:2022pth}. For fixed values of $m_a$ and $v_{\rm max}$, the blue contours are computed by plugging in $1/f_a$ from~\eqnref{fa} into~\eqnref{Gagg} for $g_{a\gamma \gamma}$, and using~\eqnref{thetaeff} to fix $\bar{\theta}_{\rm eff}$ as a function of $k$, $m_a$ and $v_{\rm max}$. We also use~\eqnref{thetaeff} to enforce the nEDM constraint $|\bar{\theta}_{\rm eff}| \leq 10^{-10}$.  Choosing a specific $k$ denoted by a red dotted line, we can access lighter axion masses along a given blue contour of fixed $v_{\rm max}$ up to the intersection point.  Explicitly, accessing smaller axion masses requires a small $k$ and hence tuning $\bar{\theta}$ closer to $\pi$.

As~\eqnref{fa} suggests for sufficiently small $m_a$ (i.e., $m_a \ll \left|\Lambda_{\text{QCD}}^2 \cos(\bar \theta- \bar \theta_{\text{eff}})/(2 N_g v_{\text{max}})\right|$), $1/f_a$ becomes insensitive to $m_a$, and approaches $1/f_a  \simeq \left| \cos(\bar \theta- \bar \theta_{\text{eff}})/(N_g v_{\text{max}} )\right|$. Physically, the vev shift from the $B_\mu$ term begins to dominate in this regime; corresponding to the kink in the blue lines of ~\figref{gagg} upon their intersection with the $k=1$ line.  In the $v_a\lesssim v_\text{max}$ region we can evaluate the corresponding magnitude of $B_\mu$ by
\begin{equation}
    |B_\mu|\lesssim \frac{\Lambda_\text{QCD}^4}{vv_a}\sqrt{1+\delta^2} = \frac{2\Lambda_\text{QCD}^4}{\sin(2\phi) v^2}\, ,
\end{equation}
with $\phi$ being the mixing angle between $v_1$ and $v_2$, leading to an estimate of $|B_\mu|\sin(2\phi) \lesssim 10^{-9}$~GeV$^2$.

In~\figref{gagg} we have enforced $\bar \theta \in \left[ \frac{\pi}{2}, \pi\right]$ leading to light Anarchic Axion masses populating the region to the left of the DFSZ band.  Note that heavier masses can populate the region to the right of the DFSZ line for $\bar{\theta} < \pi/2$. We leave the details of the heavy Anarchic Axion to future work~\cite{Elahietal2023}.

\mysec{The Quality Problem}
\begin{figure}[t!]
\includegraphics[width=0.46\textwidth]{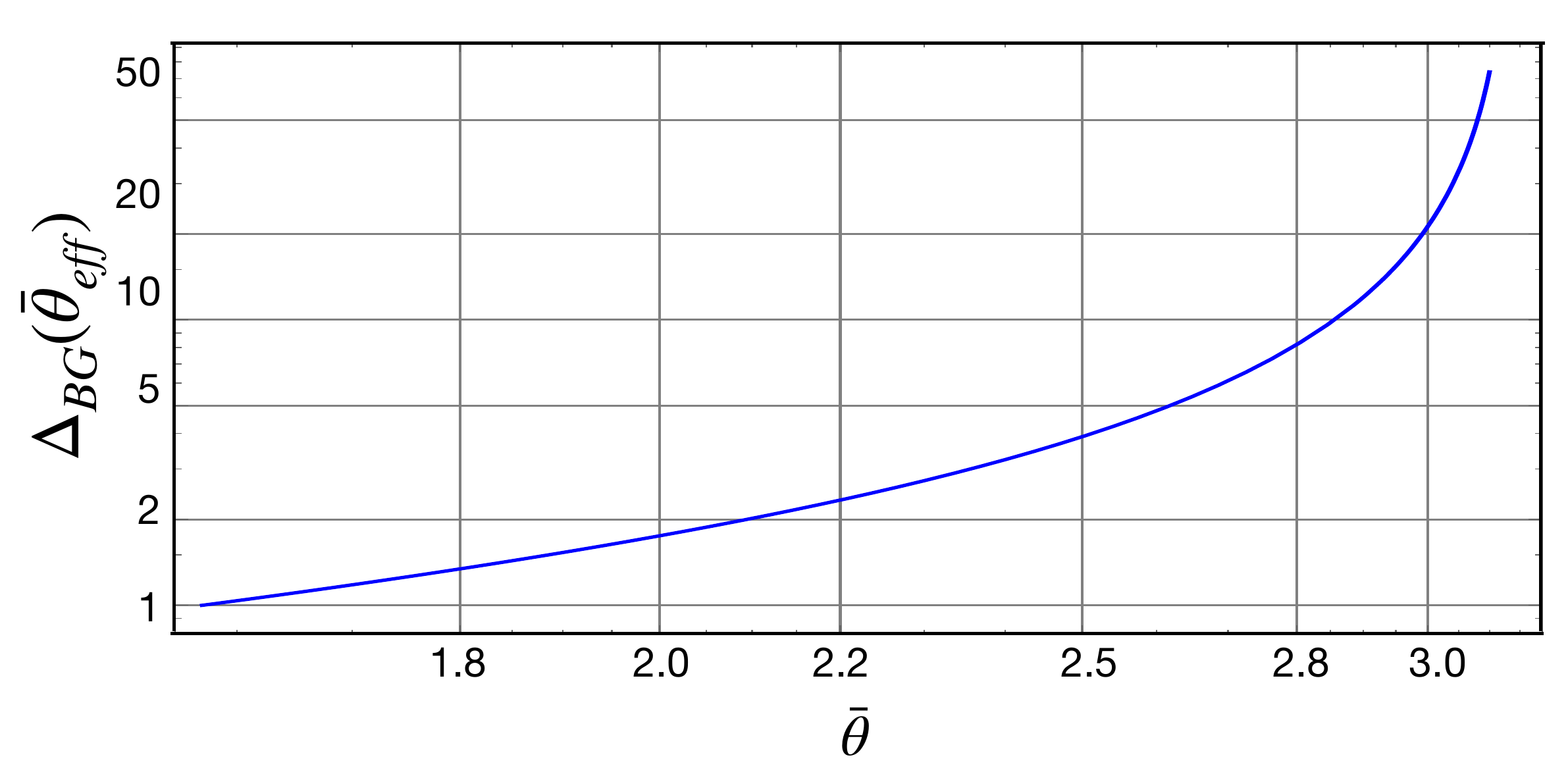}
\caption{ A measure of tuning for the variable $\bar \theta$ consistent with nEDM constraints on $\bar \theta_{\rm eff}$ given in~\eqnref{thetaeff}, fixing $m_a = 10^{-6}$~eV and $v_{\text{max}} = 10^7$~GeV.  As the figure demonstrates, the tuning exponentially increases as $\bar \theta \to \pi$, and this tuning is negligibly sensitive to alternate $m_a$ and $v_{\text{max}}$ values.}
\label{fig:finetuning}
\end{figure}
All axion models suffer from a high scale quality problem, {\it i.e.} higher dimensional operators, which are generally present from gravity effects, break the PQ symmetry at high scales and shift the axion vev. 

The quality problem of the Anarchic Axion manifests as an alignment of $\bar{\theta}$ with $\pi$ once the soft breaking $B_{\mu}$ term starts to dominate the vev shift, as is evident from the low $m_a$ plateau in~\figref{gagg}. To quantify the quality problem, we introduce a measure $\Delta_{\rm BG}$ of fine tuning, following~\cite{Ellis:1986yg, Barbieri:1987fn}:
\begin{align}
\Delta_{\rm BG}(\bar{\theta}_{\text{eff}}) \equiv \Bigl| \frac{\bar{\theta}}{\bar{\theta}_{\rm eff}} \frac{\partial \bar{\theta}_{\rm eff}}{\partial \bar{\theta}} \Bigr| \,.
\end{align}
Note that a large value of $\Delta_{\rm BG}$ implies a large tuning. Indeed, in~\figref{finetuning} we observe that $\Delta_{\rm BG}$ grows as $\bar{\theta} \rightarrow \pi$. We emphasize that the region where the Anarchic Axion exhibits a lighter than usual axion is characterized by a non-trivial fine tuning.  This approach of \emph{quantifying} the quality problem can be potentially applied to other axion models where the PQ symmetry breaking enters explicitly at high scales.  Note that this is only possible since the residual CPV $\bar{\theta}_{\text{eff}}$ is calculable.  

We also mention the possibility that the soft breaking $B_{\mu}$ term of the Anarchic Axion model can arise as the leading low energy operator matched to a Planck-suppressed PQ breaking term in canonical axion models.  We will reserve a study of the matching requirements of soft PQ breaking terms in high-quality axion models for the future.

\mysec{Discussion}   
In this Letter we have introduced the Anarchic Axion model which solves the QCD Strong CP problem parametrically while populating new regions of parameter space.  We close by mentioning that these regions with fine tuning can be motivated by, {\it e.g.} a clockwork-like ultraviolet model~\cite{Choi:2015fiu, Giudice:2016yja, Ahmed:2016viu}.  For instance, consider a $U(1)_{\text{PQ}}$ bulk gauge symmetry in a 5D warped geometry with a bulk electroweak singlet scalar, where PQ charges will be carried by a brane-localized Higgs and right-handed up-type quarks.  
Given an appropriate choice of boundary conditions for the bulk field, the Anarchic Axion model then arises as an effective description with a global $U(1)_{\text{PQ}}$ symmetry. A discrete $\mathbb{Z}_5$ symmetry can be identified with a remnant of the bulk gauge symmetry.  We leave the details of such UV completions to future work~\cite{Elahietal2023}.

While the interesting phenomenology of the Anarchic Axion model discussed in this Letter arose due to a soft PQ breaking term, other variations can produce similar phenomenology. For instance, replacing the soft breaking $B_{\mu}$ term with a $\Phi^3$ term would result in a potential which is protected by an accidental and global $\mathbb{Z}_3$ symmetry. We leave the exploration of variants of Anarchic Axion models to future work~\cite{Elahietal2023}.
We expect the cosmological production of the Anarchic Axion to proceed through a variation of the canonical misalignment mechanisms and also leave the detailed implications of Anarchic Axion dark matter to future study.

The experimental observation of an exceptionally light axion deviating from the canonical QCD axion band would be evidence for an Anarchic Axion solution to the Strong CP problem.

\mysec{Note added:}
When this work was in preparation, a related preprint appeared~\cite{Qiu:2023los}, focusing on generating high-quality axion solutions by suppressing Planck-scale operators using chiral gauged $U(1)$ symmetries.

\begin{acknowledgments}
We thank Prisco Lo Chiatto for useful conversations. We thank Raymond Co, Joshua Eby, and Alfredo Walter Mario Guerrera for useful discussions and comments on the draft.  F.E. and G.E. are grateful to CERN for their hospitality. 
This work is supported by the Cluster of Excellence {\em Precision Physics, Fundamental Interactions and Structure of Matter\/} (PRISMA${}^+$ -- EXC~2118/1) within the German Excellence Strategy (project ID 39083149). FE is also funded by grant 05H18UMCA1 of the German Federal Ministry for Education and Research (BMBF). 
\end{acknowledgments}

\bibliography{AnarchicAxion_arXiv_v2}

\onecolumngrid
\newpage
\begin{center}
  \textbf{\large Supplemental Material for: A Lighter QCD Axion from Anarchy}\\[.2cm]
  \vspace{0.05in}
  {Fatemeh Elahi, \ Gilly Elor, \ Alexey Kivel,\ Julien Laux, \ Saereh Najjari \ and \ Felix Yu}
\end{center}

\setcounter{equation}{0}
\setcounter{figure}{0}
\setcounter{page}{1}
\makeatletter
\renewcommand{\theequation}{S\arabic{equation}}
\renewcommand{\thefigure}{S\arabic{figure}}

In this supplementary material,  we first present the details of the Goldstone basis necessary for identifying the mass eigenstatates of the Anarchic Axion $a$ and the heavy field $A$. We then present the details of the derivation of $\bar \theta$. 

\section{Goldstone Basis}
\label{sec:GoldstoneBasis}

In order to identify the corresponding Goldstone bosons to the spontaneously broken $U(1)$ symmetries of hypercharge $U(1)_Y$, Peccei-Quinn $U(1)_X$ and an orthogonal global $U(1)_Z$, we perform an $O(3)$ basis rotation on the initial $U(1)_{H_1} \times U(1)_{H_2} \times U(1)_\Phi $ symmetry.  We call the Goldstone bosons $G$ for hypercharge, $a$ for Peccei-Quinn, $A$ for $U(1)_Z$ and the corresponding vevs $v$, $v_a$ and $v_A$, respectively. The rotation matrix acting on the unphysical angular modes $a_1$, $a_2$ and $a_3$ of $H_1$, $H_2$ and $\Phi$ respectively, reads
\begin{align}
    \begin{pmatrix}
    G\\a\\A
    \end{pmatrix}=
    \begin{pmatrix}
    s_\phi c_\gamma &-c_\phi c_\gamma &-s_\gamma \\
    c_\phi c_\beta -s_\phi s_\beta s_\gamma &s_\phi c_\beta +c_\phi s_\beta s_\gamma &-s_\beta c_\gamma \\
    c_\phi s_\beta +s_\phi c_\beta s_\gamma &s_\phi s_\beta -c_\phi c_\beta s_\gamma &c_\beta c_\gamma
    \end{pmatrix}
    \begin{pmatrix}
    a_1\\a_2\\a_3
    \end{pmatrix},
    \label{eqn:mixingmatrix}
\end{align}
with $\tan\phi =v_1/v_2$, $\tan\beta=v_A/v_a$ and $\gamma =0$, since $\Phi$ is not charged under $U(1)_Y$ and therefore $a_3$ does not mix into $G$. The corresponding vev relations derived from orthogonality conditions are 
\begin{equation}
\label{eqn:vevs}
    v_1=v\sin{\phi}\, , \quad v_2=v\cos{\phi}\, , \quad v_3=v_a\sin{\beta}=v_A\cos{\beta}\, , \quad v_a\cos\beta=v\sin\phi\cos\phi \ .
\end{equation}
Under this basis rotation the angular potential in our model transforms as
\begin{align}
    V_\text{ang} =& - |B_\mu|  \Bigl[\prod_{i = 1}^2\left(v_i + h_i \right) \Bigr] \, \cos \left(\sum_{i=1}^2\frac{a_i}{v_i}  - \theta_\mu \right)  - \frac{|C_\lambda| }{\sqrt{2}} \Bigl[\prod_{i = 1}^3\left(v_i + h_i \right) \Bigr] \cos \left(\sum_{i=1}^3\frac{a_i}{v_i} - \theta_\lambda \right) \nonumber \\
    \overset{O(3)}{\longrightarrow} & -|B_\mu| \Bigl[\prod_{i = 1}^2\left(v_i + h_i \right) \Bigr]\cos(\frac{a}{v_a}+\frac{A}{v_A}\tan^2 \beta -\theta_\mu)
    -\frac{|C_\lambda|}{\sqrt{2}} \Bigl[\prod_{i = 1}^3\left(v_i + h_i \right) \Bigr]\cos(\frac{A}{v_A}\sec^2 \beta -\theta_\lambda) \ . \label{eqn:scalarPotential}
\end{align}
In this new basis the dependence of the angular potential on $G$ and $\phi$ conveniently drops out. We can replace $\tan\beta$ by a parameter $\delta\equiv v_A/v_a$, leading to
\begin{equation}
    \tan^2\beta=\delta^2\, , \quad \sec^2\beta=1+\delta^2\, , \quad v_1v_2=\frac{vv_a}{\sqrt{1+\delta^2}}\, , \quad v_3=\frac{v_A}{\sqrt{1+\delta^2}}\, .
\end{equation}
Electroweak constraints on the $A$ pseudoscalar state will generally require that $v_A \gtrsim v$.

\section{Derivation of $\bar{\theta}$}
We present the derivation of the Anarchic Axion strong CP-violating parameter $\bar{\theta}$ and discuss its relaxation. The overall observable strong CP-violating is  basis independent and must be defined by a unique linear combination of the CP-violating phases, similar to the SM, where the quark phases and $\theta_{\text{QCD}}$ contribute to the unique observable $\bar{\theta}_{\text{SM}}$. 
The CPV of the Anarchic Axion can be reshuffled by applying a $U(1)$ transformation on the fields $H_1$, $H_2$, $\Phi$ and the SM quarks, thereby redistributing the CP-violating phases $\theta_\mu$, $\theta_\lambda$, and $\bar{\theta}_{\text{SM}}$. Starting with the Lagrangian in~\eqnref{general_potential}, the phases can be rotated into the quark masses by applying the transformations
\begin{align}
    H_2 \rightarrow e^{i\theta_\mu} H_2 \ , \quad 
    \Phi \rightarrow e^{i(-\theta_\mu+\theta_\lambda)} \Phi \ . \label{eqn:derThetaBarTranf}
\end{align}
This introduces real prefactors in all terms of \eqnref{general_potential} and an additional phase factor $e^{i\theta_\mu}$ in the SM Yukawa couplings 
\begin{align}
   \bar{Q}_L Y_d H_2 d_R \rightarrow & \bar{Q}_L Y_d H_2 e^{i\theta_\mu} d_R \equiv \bar{Q}_L (Y_d') H_2 d_R \ ,
\end{align}
where $Q_L$ are the left-handed quark doublets, $d_R$ are the right-handed down-type quarks, and $Y_d'$ is the new Yukawa coupling matrix. Note that the unphysical $\theta_\lambda$ phase is absorbed by $\Phi$. The resulting strong CPV is defined through
\begin{align}
    (\theta_{\text{QCD}} - \arg(\det(Y_d' Y_u)))\ G\tilde{G} & = (\theta_{\text{QCD}}  - \arg(\det(Y_d Y_u))) - N_g \theta_\mu)\ G\tilde{G} \nonumber \\
    & = (\bar{\theta}_{\text{SM}} - N_g \theta_\mu)\ G\tilde{G} \equiv N_g \bar{\theta} \ G\tilde{G} \, . \label{eqn:definitionThetaBar}
\end{align}
Note that one can likewise choose to transform $H_1$ in~\eqnref{derThetaBarTranf} such that the phase factor appears in $Y_u$, resulting in an identical $\bar{\theta}$.

\end{document}